\documentclass[a4paper,11pt]{article}
\pdfoutput=1

\usepackage{jinstpub}

\usepackage{lineno}
\modulolinenumbers[5]
\usepackage{siunitx}
\usepackage{subcaption}
\usepackage{url}
\usepackage{color, soul}

\title{\boldmath Towards MPGDs with embedded pixel ASICs}

\author[a,1]{L. Scharenberg,\note{Corresponding author.}}
\author[a]{J. Alozy,}
\author[a]{W. Billereau,}
\author[a]{F. Brunbauer,}
\author[a]{M. Campbell,}
\author[a]{P. Carbonez,}
\author[a,b]{K. Fl{\"o}thner,}
\author[c]{F. Garcia,}
\author[a,d]{A. Garcia-Tejedor,}
\author[a,d]{T. Genetay,}
\author[e]{K. Heijhoff,}
\author[a]{D. Janssens,}
\author[a]{S. Kaufmann,}
\author[a]{M. Lisowska,}
\author[a]{X. Llopart,}
\author[a]{M. Mager,}
\author[a]{B. Mehl,}
\author[f,a]{H. Muller,}
\author[a]{R. de Oliveira,}
\author[a]{E. Oliveri,}
\author[a,g]{G. Orlandini,}
\author[h,a]{D. Pfeiffer,}
\author[a]{F. Piernas Diaz,}
\author[a]{A. Rodrigues,}
\author[a]{L. Ropelewski,}
\author[h]{J. Samarati,}
\author[e]{M. van Beuzekom,}
\author[a]{M. van Stenis,}
\author[a]{R. Veenhof,}
\author[i]{and M. Vicente}

\affiliation[a]{European Organization for Nuclear Research (CERN), 1211 Geneva 23, Switzerland}
\affiliation[b]{Helmholtz-Institut f{\"u}r Strahlen- und Kernphysik, University of Bonn, Nu{\ss{}}allee 14-16, 53115 Bonn, Germany}
\affiliation[c]{Helsinki Institute of Physics (HIP), P.O. Box 64, FI-00014 University of Helsinki, Finland}
\affiliation[d]{Institute of Radiation Physics, Centre hospitalier universitaire vaudois (CHUV), Rue du Grand-Pr\'{e} 1, 1007 Lausanne, Switzerland}
\affiliation[e]{Nikhef, Science Park 105, 1098 XG Amsterdam, the Netherlands}
\affiliation[f]{Physikalisches Institut, University of Bonn, Nu{\ss{}}allee 12, 53115 Bonn, Germany}
\affiliation[g]{Friedrich-Alexander-Universit{\"a}t Erlangen-N{\"u}rnberg, Schlo\ss{}platz 4, 91054 Erlangen, Germany}
\affiliation[h]{European Spallation Source ERIC (ESS), Box 176, SE-221 00 Lund, Sweden}
\affiliation[i]{D\'{e}partement de physique nucl\'{e}aire et corpusculaire, University of Geneva, 24 quai Ernest-Ansermet, 1205 Gen\`{e}ve 4, Switzerland}

\emailAdd{lucian.scharenberg@cern.ch}

\abstract{Combining gaseous detectors with a high-granularity pixelated charge readout enables experimental applications which otherwise could not be achieved.
This includes high-resolution tracking of low-energetic particles, requiring ultra-low material budget, X-ray polarimetry at low energies ($\lessapprox \SI{2}{keV}$) or rare-event searches which profit from event selection based on geometrical parameters.
In this article, the idea of embedding a pixel ASIC --- specifically the Timepix4 --- into a micro-pattern gaseous amplification stage is illustrated.
Furthermore, the first results of reading out a triple-GEM detector with the Timepix4 (GEMPix4) are shown, including the first X-ray images taken with a Timepix4 utilising Through Silicon Vias (TSVs).
Lastly, a new readout concept is presented, called the `Silicon Readout Board', extending the use of pixel ASICs to read out gaseous detectors to a wider range of HEP applications.}


\proceeding{8\textsuperscript{th} International Conference on Micro-Pattern Gaseous Detectors\\
  14--18 October 2024\\
  Hefei, China}

\begin{document}
\maketitle
\flushbottom

%\linenumbers

\section*{Introduction}
In this article, a new research line within the Work Package 2 (Gas Detectors) of CERN's EP R\&D programme \cite{ep-rnd-1,ep-rnd-2} is presented.
The goal is to investigate the charge readout of gaseous detectors with high-granularity hybrid pixel Application-Specific Integrated Circuits (ASICs).
The ASIC of choice for this is the Timepix4 \cite{timepix4}, where instead of bump-bonding a semiconductor sensor to it, the bump-bond pads are used as charge collection pads.
This approach for reading out MPGDs was successfully applied to experiments with previous versions of the Timepix in the GridPix \cite{gridpix} and the GEMPix \cite{gempix} detectors.
With the Timepix4 however, more and improved features are available, such as a larger area (around $\SI{7}{cm^2}$, with $\num{448}\times\SI{512}{pixels}$ with $\SI{55}{\micro m}$ square pitch), a time resolution of $\SI{60}{ps}$ and hit rates of up to $\SI{3}{MHz/mm^2}$ in data-driven readout mode.
Most importantly, the Timepix4 contains Through Silicon Vias (TSVs), which enable a full back-side connection of the ASIC to the back-end electronics, making wire bonds obsolete.
This is not only advantageous in covering large areas without any dead region in between Timepix4 ASICs, but it is also the only way a pixelated front-end ASIC can be embedded into a micro-pattern gaseous amplification stage.

\section{Embedding concept}
\begin{figure}[t]
    \centering
    \begin{subfigure}{86.8125mm}
        \centering
        \includegraphics[width = \columnwidth]{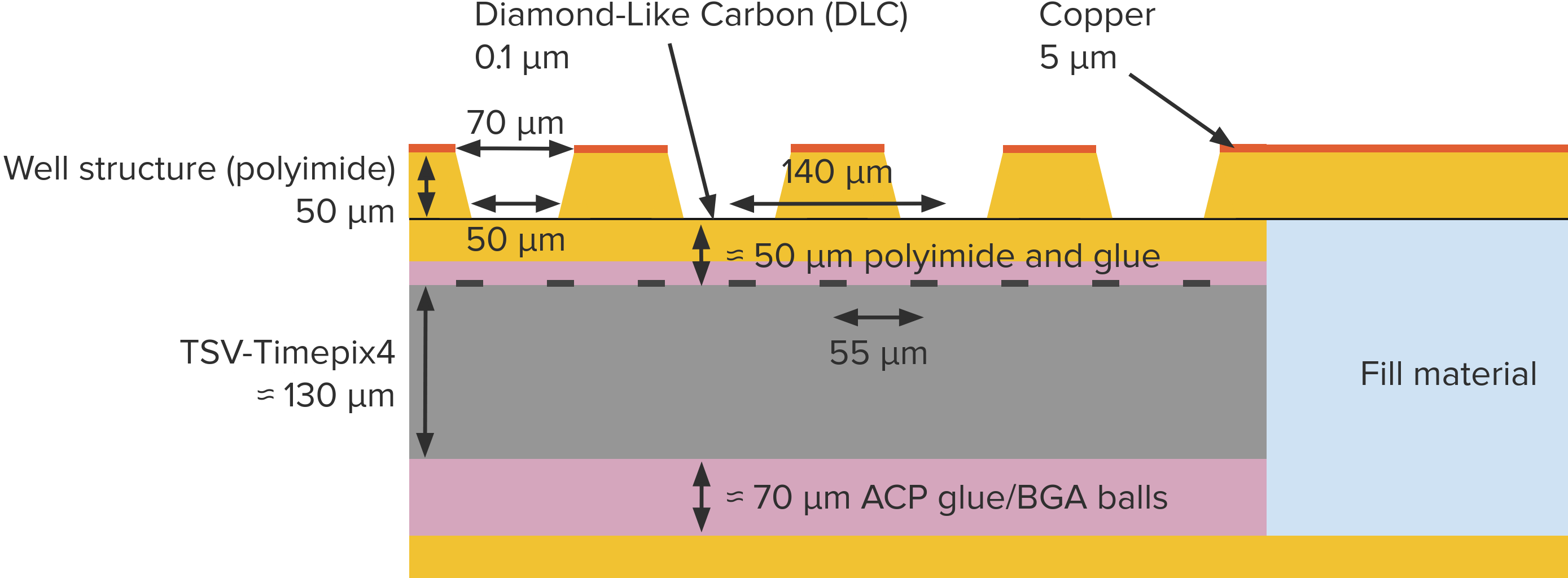}
        \caption{}
        \label{fig:embedding-concept}
    \end{subfigure}
    \hspace{0.1\columnwidth}
    \begin{subfigure}{33.288590604mm}
        \centering
        \includegraphics[width = \columnwidth]{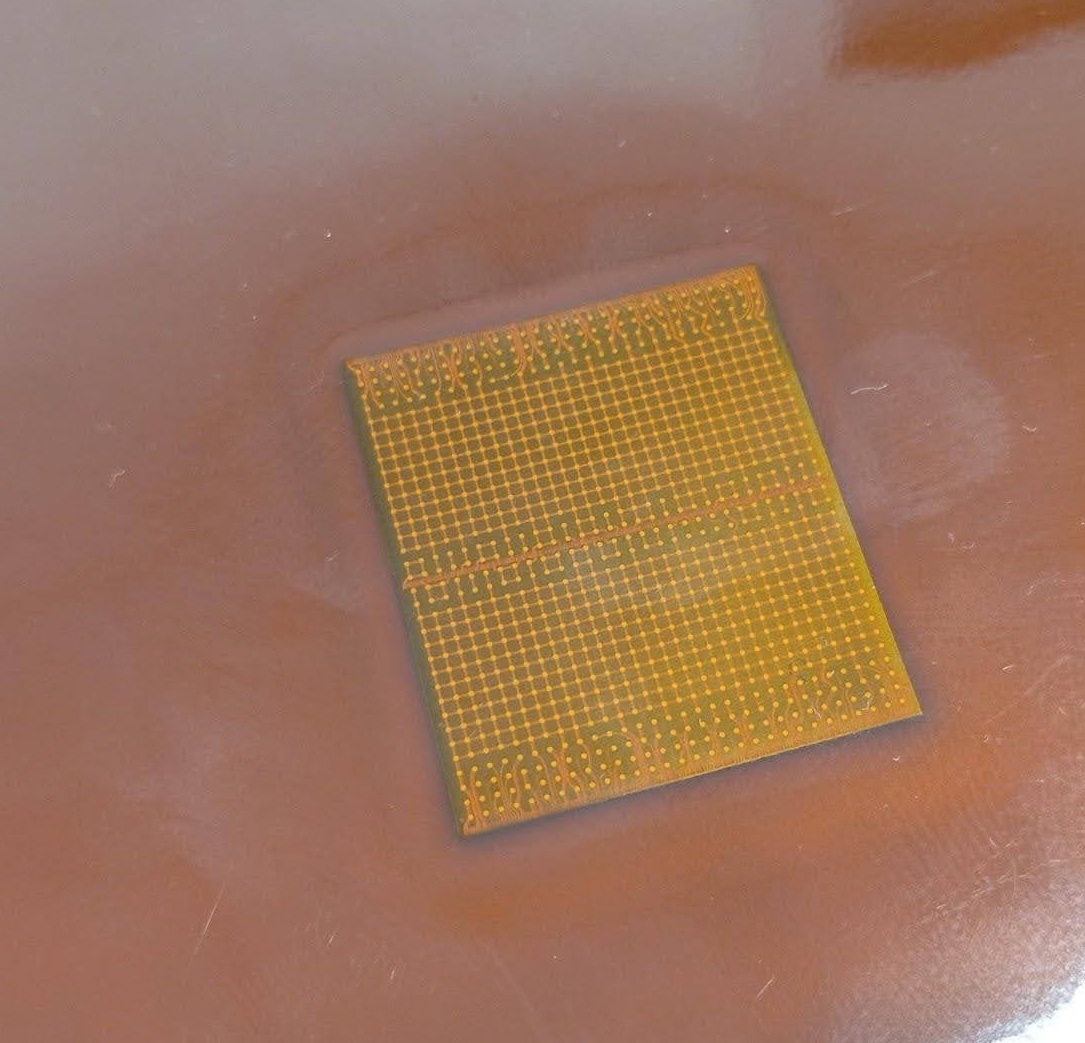}
        \caption{}
        \label{fig:embedding-photograph}
    \end{subfigure}
    \caption{(a) Sketch of the embedding concept with the TSV-Timepix4.
        (b) Photograph of the first mechanical embedding result. The Timepix4 has been laminated between two polyimide foils with additional fill material in between the foils to account for its $\approx\SI{130}{\micro m}$ height.}
    \label{fig:embedding}
\end{figure}

The concept of embedding the Timepix4 into a gaseous amplification stage, in particular a \textmu{}RWELL \cite{urwell}, is inspired by a presentation given at the 7\textsuperscript{th} MPGD conference in 2022 \cite{embedding} and the so-called `MAPS foil' \cite{maps-foil}.
The goal is to use standard micro-pattern technologies to laminate the Timepix4 into a polyimide flex-PCB, with the connection to the readout system on one side and the amplification stage on the other side (Fig.~\ref{fig:embedding}).
First steps to investigate the feasibility of this have been undertaken:
\begin{itemize}
    \item The signal induction from the \textmu{}RWELL amplification holes to the Timepix4 pixels was studied with simulations.
        Due to the high granularity of the Timepix with respect to the structure size of the \textmu{}RWELL, no relevant loss of signal is observed.
    \item \textmu{}RWELL foils have been produced to evaluate their performance employing a single channel anode.
        With mixtures of Ar/CO\textsubscript{2} (70/30) and Ar/CO\textsubscript{2}/CF\textsubscript{4} (45/15/40)
        %\textbf{and Ar/iC\textsubscript{4}H\textsubscript{10} (98/2)}
        detector gains of $\num{e4}$ have been reached.
    \item Mechanical integration tests, i.e. without electrical connectivity to the Timepix4, have been successfully performed (Fig.~\ref{fig:embedding-photograph}).
\end{itemize}

\section{Development and performance evaluation with the GEMPix4}
\begin{figure}[t!]
    \centering
    \begin{subfigure}{60.011741683mm}
        \centering
        \includegraphics[width = \columnwidth]{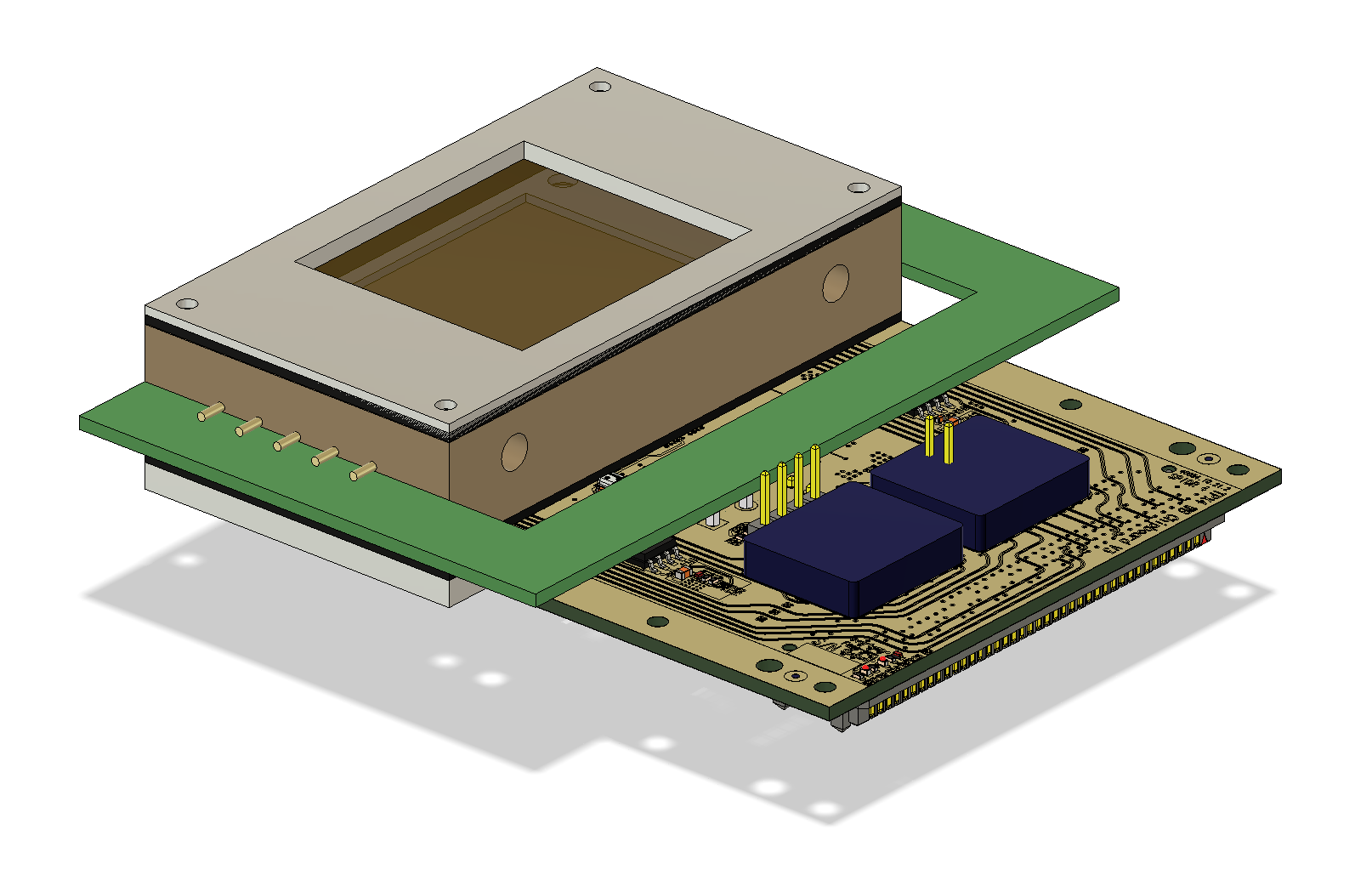}
        \caption{}
        \label{fig:detector-drawing}
    \end{subfigure}
    \hspace{0.1\columnwidth}
    \begin{subfigure}{38mm}
        \centering
        \includegraphics[width = \columnwidth]{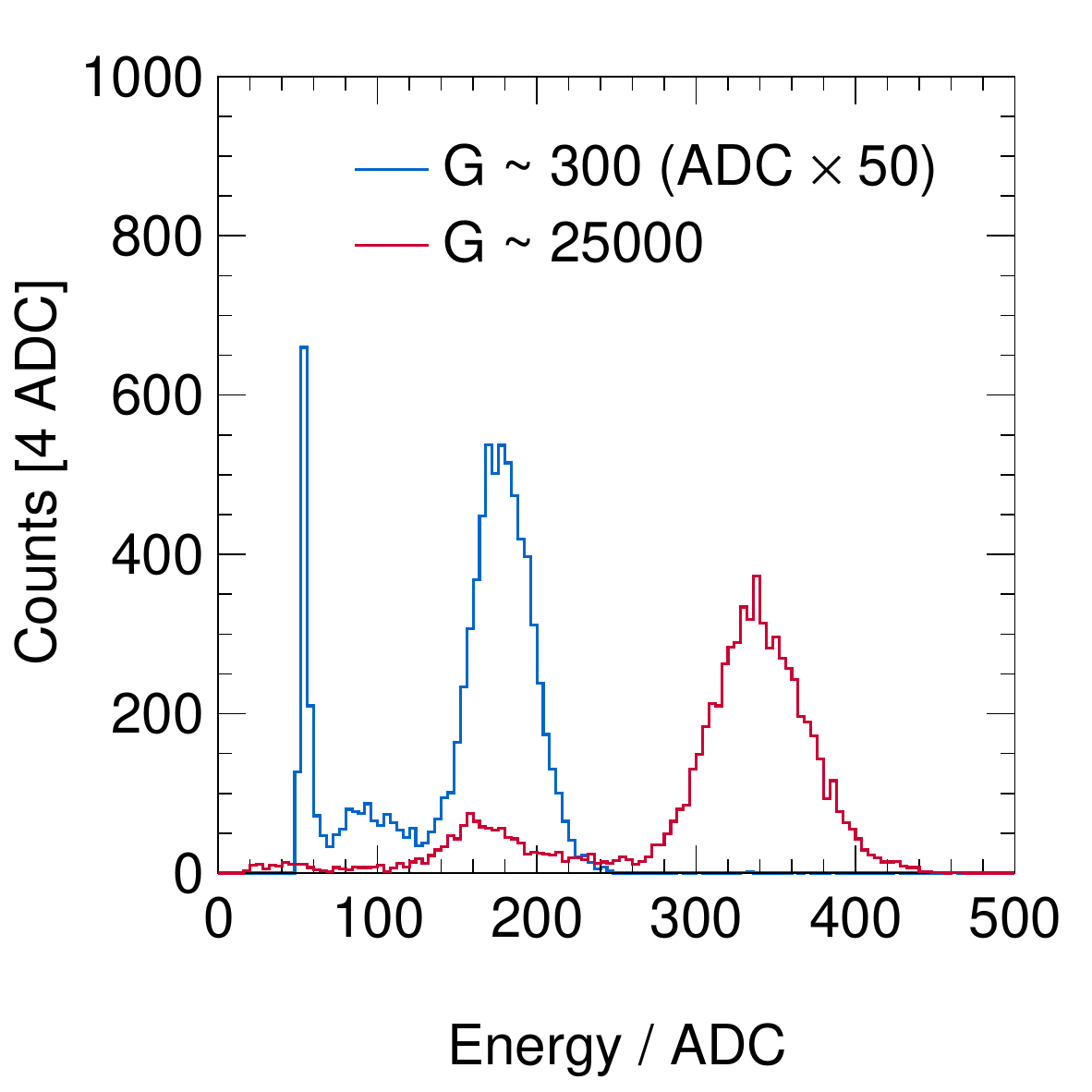}
        \caption{}
        \label{fig:detector-energy-resolution}
    \end{subfigure}
    \caption{(a) Drawing of the GEMPix4 detector, based on the SPIDR4 Nikhef carrier board.
        (b) Spectra of \textsuperscript{55}Fe at different detector gains, measured with the triple-GEM detector used for the GEMPix4, but without the Timepix4 as the anode, but a single channel copper anode.}
    \label{fig:detector}
\end{figure}

As part of the research activities, the GEMPix4 was developed (Fig.~\ref{fig:detector-drawing}), as a next version of the GEMPix detector family \cite{gempix-overview,la-gempix}.
The use of a well-known, established detector technology, allowed us to get accustomed to the combination of MPGDs with the Timepix4, including all the necessary software developments and optimisations.
However, instead of following the detector layout used in previous versions of the GEMPix, a COMPASS-like triple-GEM detector \cite{compass-gem} with a $\SI{3}{mm}$ wide drift gap, $\SI{2}{mm}$ wide transfer and induction gaps and a gas mixture of Ar/CO\textsubscript{2} (70/30) was used.
In some first tests, with a $\num{2.5}\times\SI{3.0}{cm^2}$ single-channel anode, i.e. without the Timepix4, gains of $\num{2.5e4}$ were reached, without encountering discharges --- the voltage was not further increased, i.e. the discharge limit might be even higher.
At the same time, an excellent energy resolution of $\SI{24}{\percent}$ FWHM could be obtained at low detector gains of $\num{300}$ (Fig.~\ref{fig:detector-energy-resolution}).

\begin{figure}[t!]
    \centering
    \begin{subfigure}{42.666666667mm}
        \centering
        \includegraphics[width = \columnwidth]{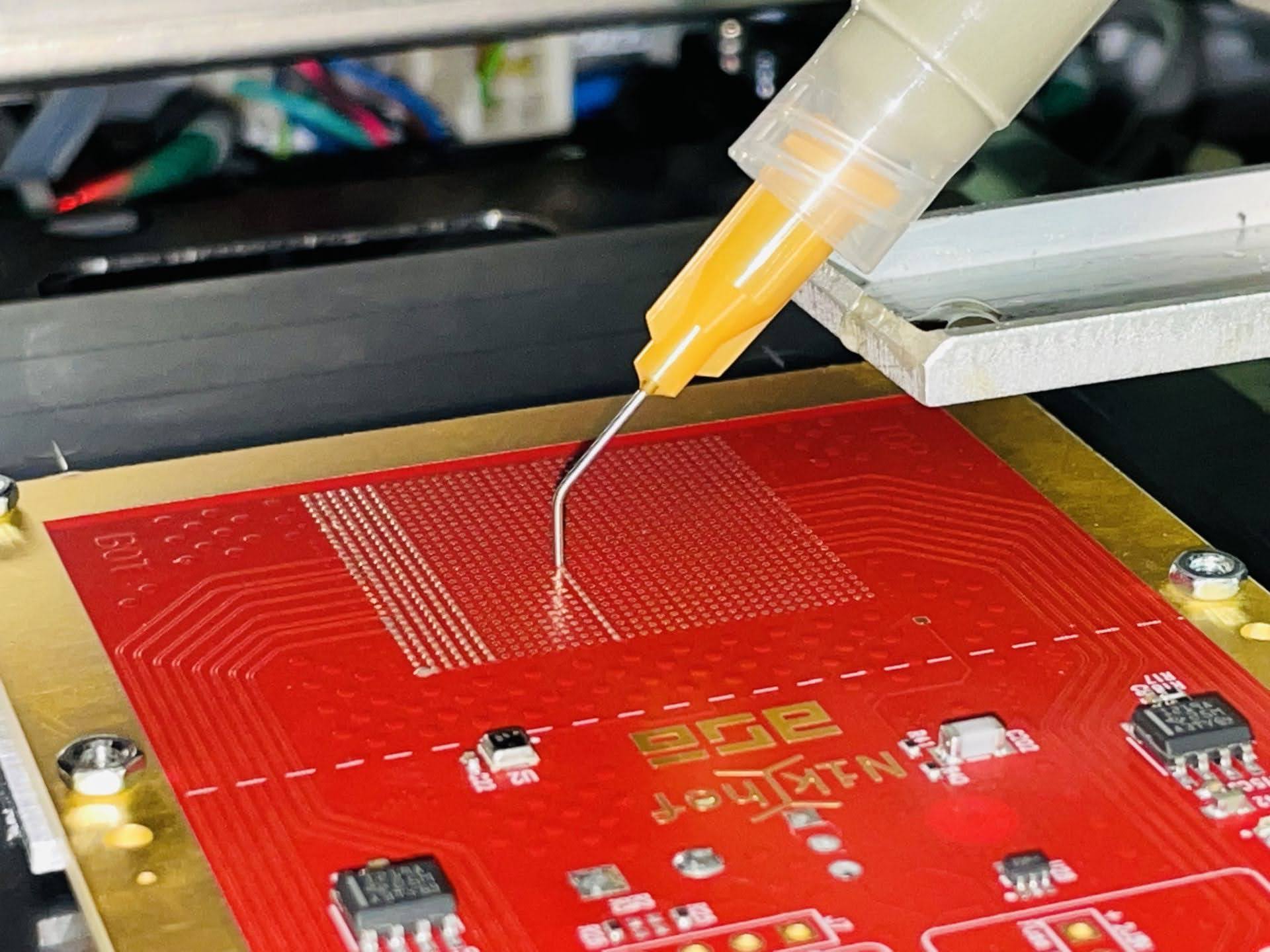}
        \caption{}
        \label{fig:assembly-acp}
    \end{subfigure}
    \hspace{0.015\columnwidth}
    \begin{subfigure}{31.843575419mm}
        \centering
        \includegraphics[width = \columnwidth]{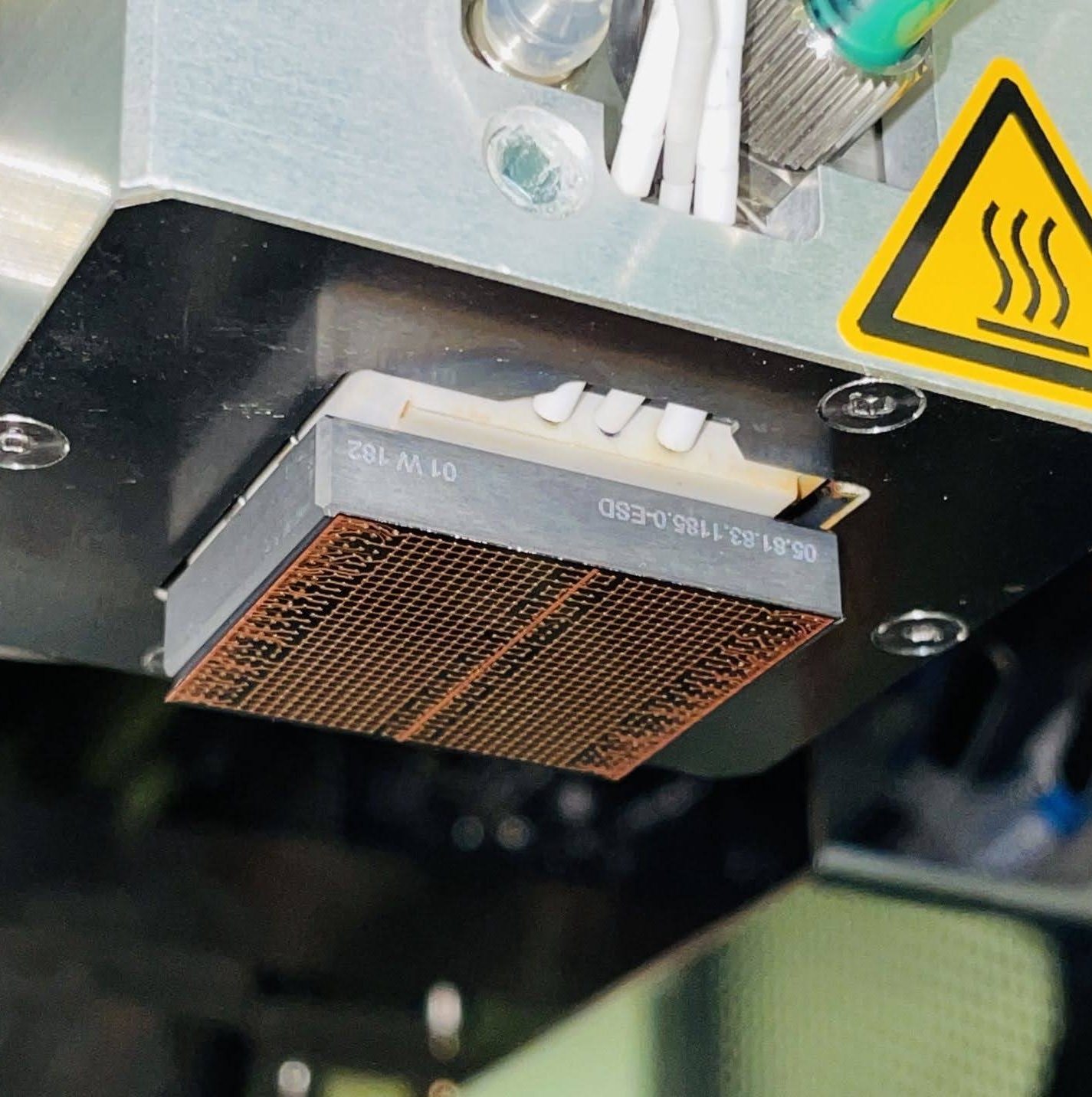}
        \caption{}
        \label{fig:assembly-flip-chip}
    \end{subfigure}
    \hspace{0.015\columnwidth}
    \begin{subfigure}{32mm}
        \centering
        \includegraphics[width = \columnwidth]{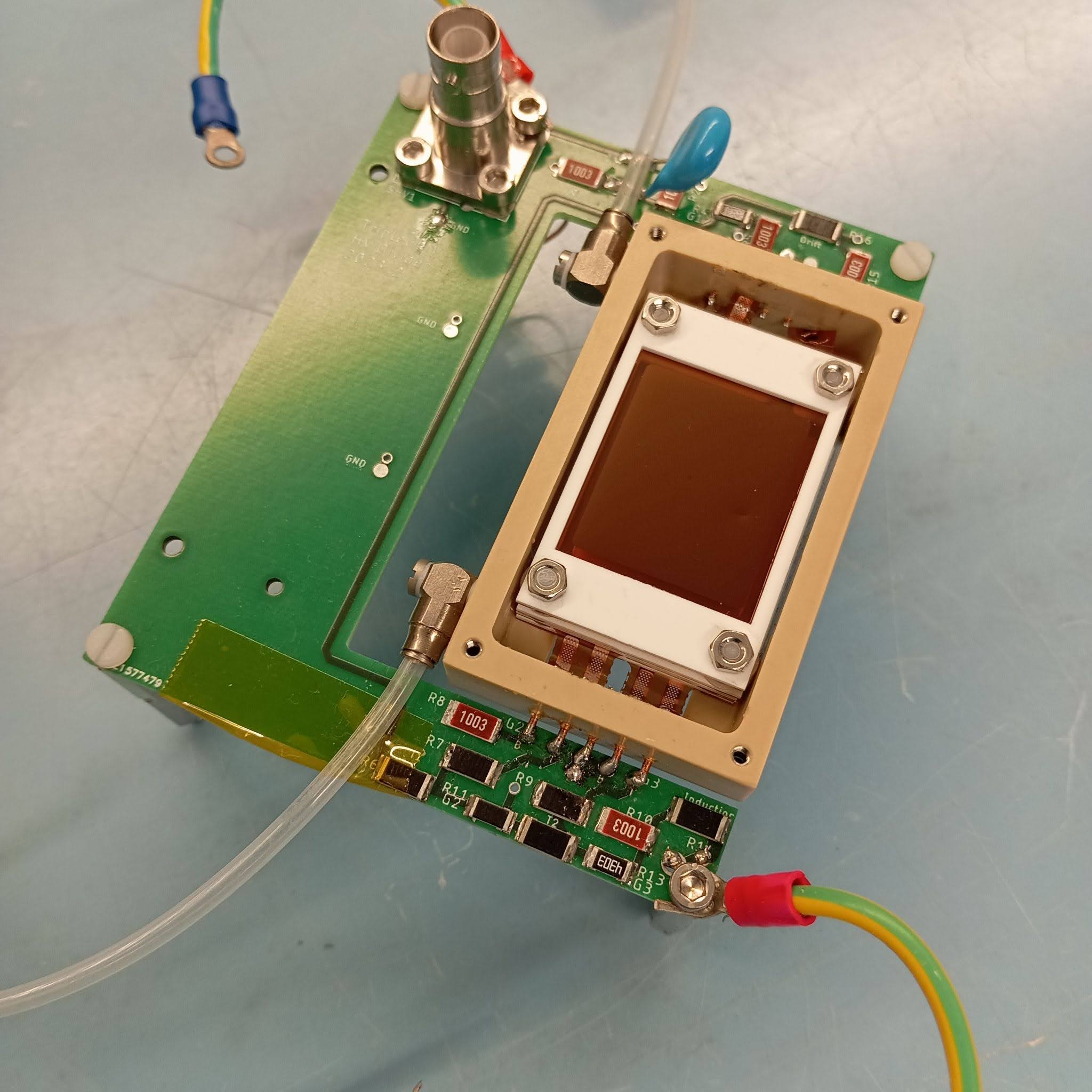}
        \caption{}
        \label{fig:assembly-detector}
    \end{subfigure}
    \hspace{0.015\columnwidth}
    \begin{subfigure}{28.309252218mm}
        \centering
        \includegraphics[width = \columnwidth]{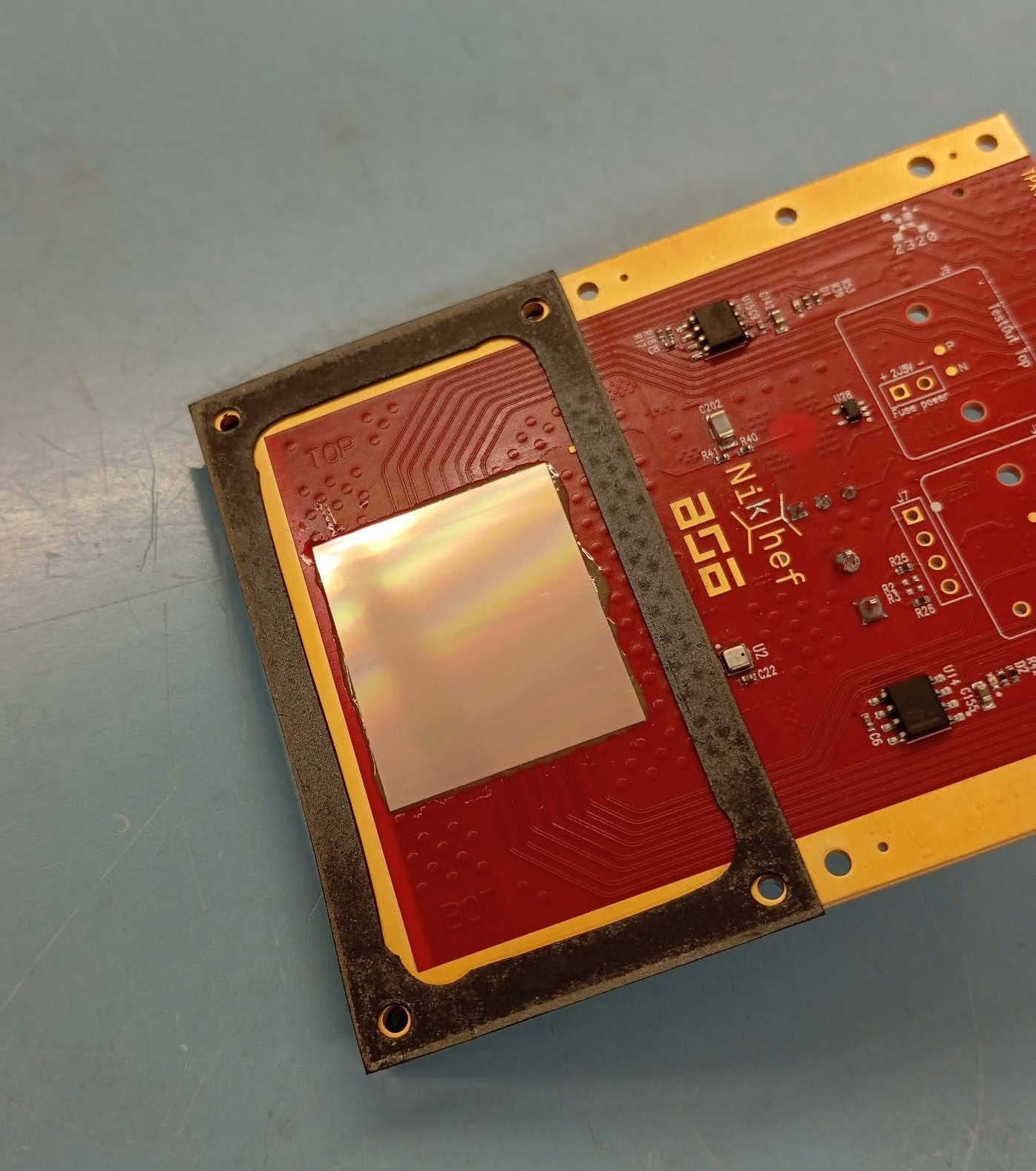}
        \caption{}
        \label{fig:assembly-carrier}
    \end{subfigure}
    \caption{Assembly and components of the GEMPix4 detector: (a) deposition of the ACP on the SPIDR4 Nikhef carrier board,
        (b) backside of the Timepix4 with the TSVs and its redistribution layer to be mounted on the carrier board,
        (c) triple-GEM stack in the opened detector housing with high-voltage divider,
        (d) TSV-Timepix4 on the carrier board with a gasket around.}
    \label{fig:assembly}
\end{figure}

\begin{figure}[t!]
    \centering
    \begin{subfigure}{42.142mm}
        \centering
        \includegraphics[width = \columnwidth]{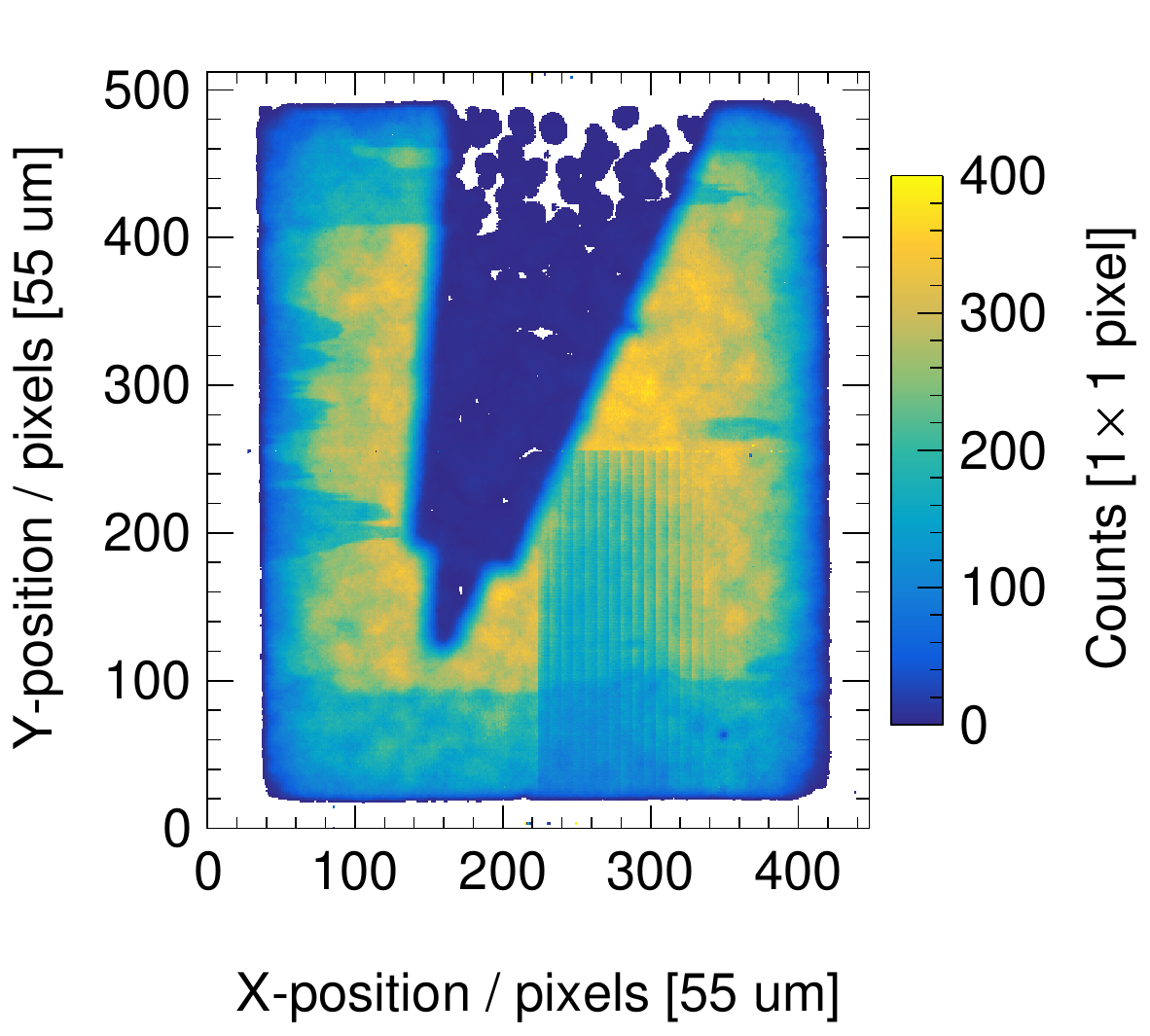}
        \caption{Image of a pen}
        \label{fig:measurements-image}
    \end{subfigure}
    \hspace{0.02\columnwidth}
    \begin{subfigure}{38mm}
        \centering
        \includegraphics[width = \columnwidth]{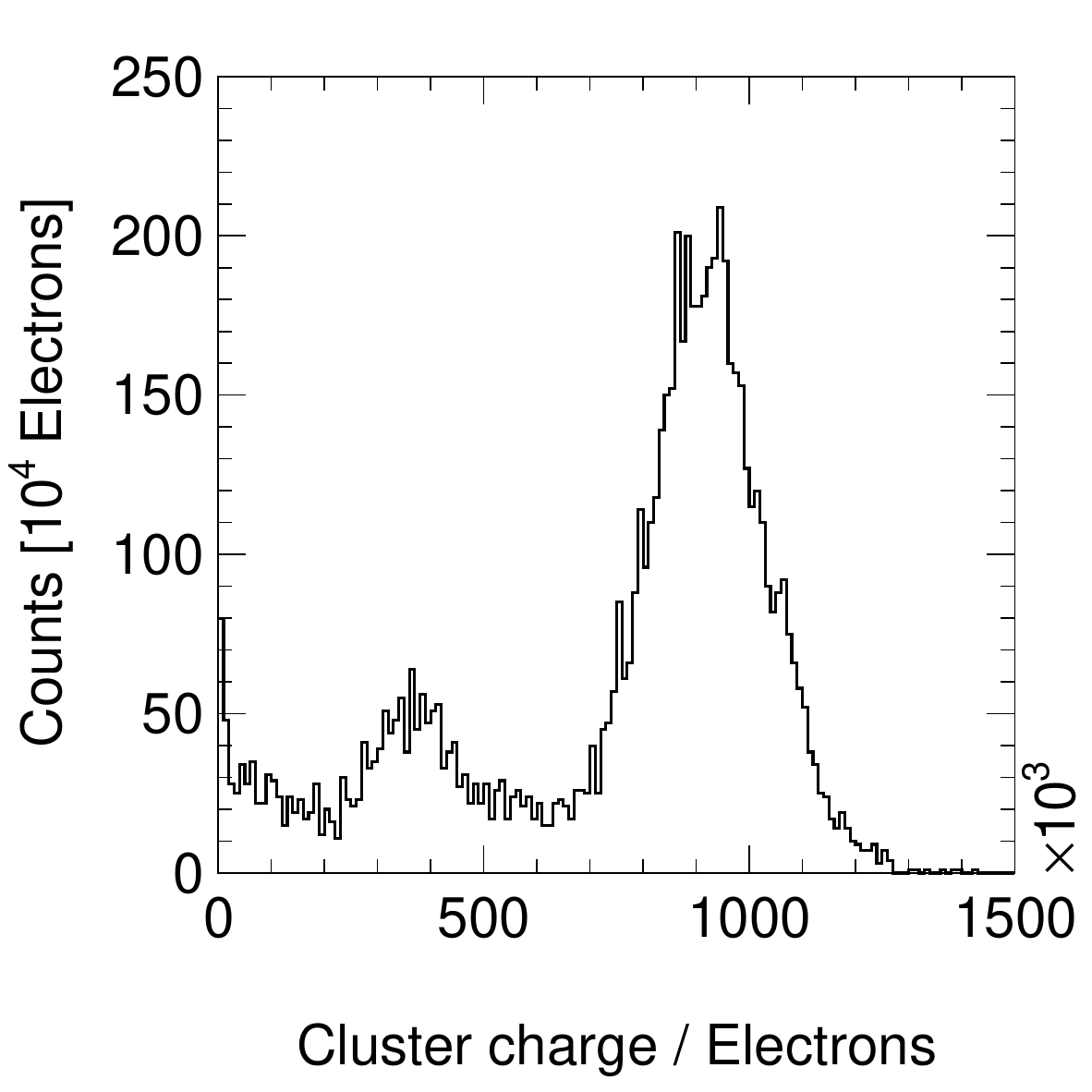}
        \caption{\textsuperscript{55}Fe spectrum}
        \label{fig:measurements-spectrum}
    \end{subfigure}
    \hspace{0.02\columnwidth}
    \begin{subfigure}{45.6mm}
        \centering
        \includegraphics[width = \columnwidth]{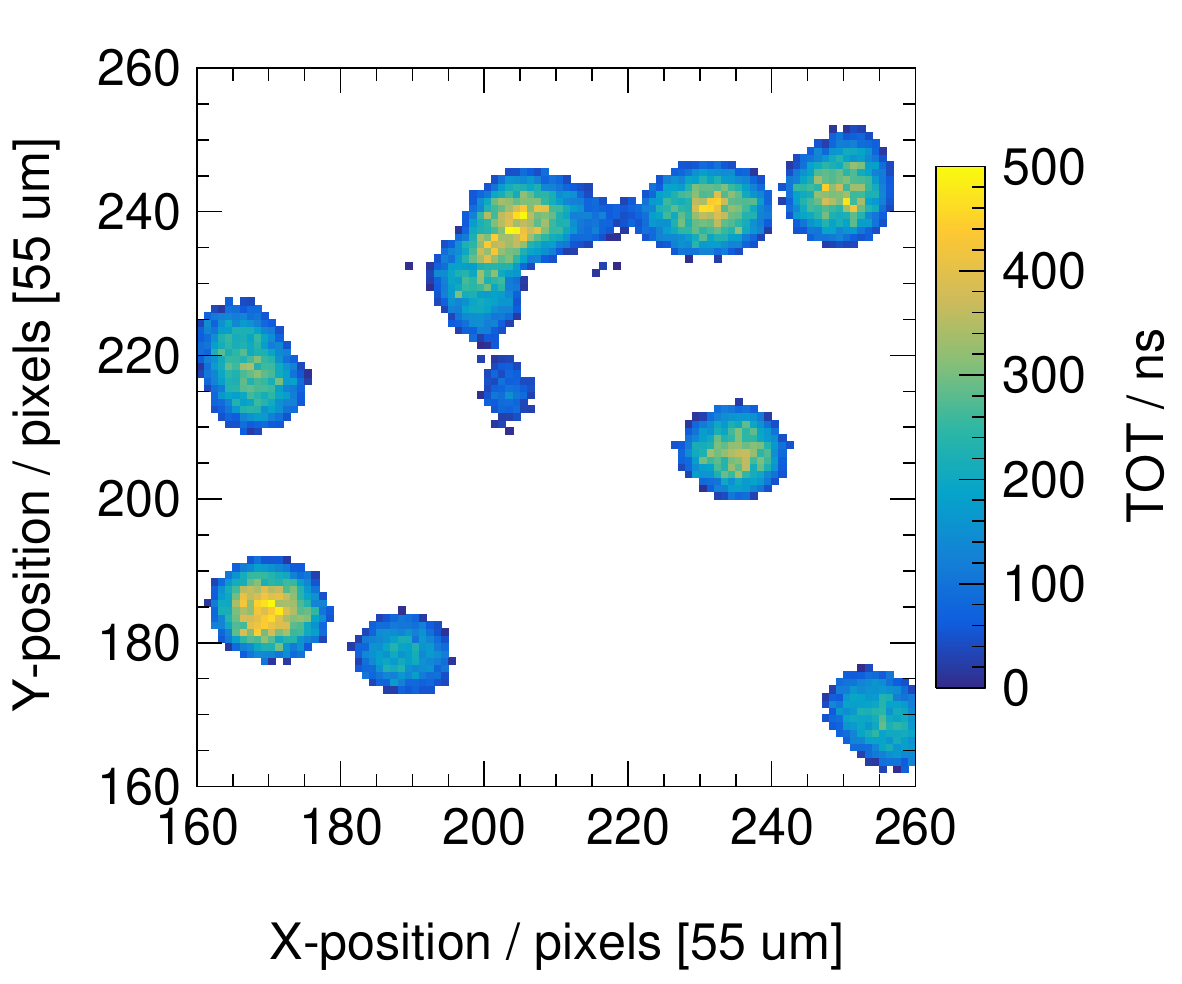}
        \caption{Individual \textsuperscript{90}Sr interactions}
        \label{fig:measurements-clusters}
    \end{subfigure}
    \caption{Results obtained with the GEMPix4 in the data-driven readout mode.}
    \label{fig:measurements}
\end{figure}

Afterwards, the single-channel anode was removed and the Timepix4 was mounted on the chip carrier board, which also serves as the base plane for the triple-GEM detector housing (Fig.~\ref{fig:assembly-detector}).
The electrical and mechanical connection for the TSV-Timepix4 is done with Anisotropic Conductive Paste (ACP).
The ASIC is read out with the SPIDR4 \cite{spidr,spidr3,spidr4} developed at Nikhef.
The first results obtained with that particular detector assembly, operated in the data-driven readout, are shown in Fig.~\ref{fig:measurements}.
In Fig.~\ref{fig:measurements-image}, the first X-ray image taken with a TSV-Timepix4 can be seen.
It shows an integrated hit map of all active pixels, with the shape of a pen, that has been placed in front of the detector window, showing a significantly lower amount of hits.
In addition, various structures can be seen.
The vertical lines in the lower-right quadrant are a consequence of bandwidth issues in the prototype Timepix4v1 ASIC, which was used for the here shown measurements; for the v2 and v3 production versions, this issue was fixed.
The other observable features on the left and right edges of the image are most likely caused by deposits on the Timepix4, as they did not appear in a second measurement with a different Timepix4.
In Fig.~\ref{fig:measurements-spectrum}, the \textsuperscript{55}Fe spectrum measured with the Timepix4 at a detector gain of around $\num{5000}$ is shown.
In Fig.~\ref{fig:measurements-clusters}, the hits from a $\SI{10}{ms}$ time slice of the continuous data stream are shown.
Each roundish structure corresponds to a single particle interaction, with the particles being emitted from a \textsuperscript{90}Sr source.
This also means that typically $\sim\num{e2}$ pixels are involved for each recorded interaction in a GEMPix detector.

\section{Future perspectives}
Although the high-granularity readout of MPGDs, e.g. with the Timepix4, offers very specific possibilities, most gaseous detector applications --- specifically HEP experiments, e.g. large-scale Time Projection Chambers (TPCs), MPGD-based calorimetry or the thousands of square meters covered by the experiments muon system --- do not require such a fine granularity.
Nonetheless, they all would profit from an ambiguity-free readout with finer granularity than at the current stage, e.g. $\num{1}\times\SI{1}{mm^2}$ pads instead of $\num{10}\times\SI{10}{mm^2}$.
One possibility to achieve this could be to scale up the embedding method and adjust it to other front-end ASICs, with larger pixel pitch.
Two examples would be the ALTIROC \cite{altiroc} and the ETROC \cite{etroc}, which both have square pixel pitches of $\SI{1.3}{mm}$.

Another possibility could be the so-called `Silicon Readout Board', which is a new idea introduced by the authors in the following.
A classical anode PCB with the front-end electronics externally connected, i.e. outside of the active detection area, would require very tedious routing and would be very expensive in terms of production, especially if pads of $\num{1}\times\SI{1}{mm^2}$ size would need to be read out --- and it does not include the costs of the required front-end electronics.
Given the typical costs of a silicon wafer, a readout structure with $\num{100}\times\SI{100}{mm^2}$ area and $\num{e4}$ readout pads could be realised, at lower costs than a comparable PCB and with the front-end electronics already included.
In addition, because the pad/pixel size is sufficiently large, compared to typical hybrid pixel ASICs, the production could be done in larger, i.e. cheaper, production processes for example $\SI{130}{nm}$.
It also would allow for additional redundancy and discharge protection on the readout board to increase the reliability of the front-end electronics.

\section*{Conclusion}
A new research line on the high-granularity readout of MPGDs has been started within CERN's EP R\&D programme.
One of the goals is to explore the possibility of embedding hybrid pixel ASICs into micro-pattern amplification stages.
Here, one profits from the capabilities of the Timepix4 ASIC with its Through Silicon Vias.
At the same time, the GEMPix4 was developed and successfully read out.
The first results are encouraging to continue this research line and optimise the detector performance.
Lastly, a new concept of a readout board for gaseous detectors has been presented.
Profiting from the capabilities of semiconductor wafer production, a fine-granularity pad readout board could be realised at improved cost and performance compared to current standard PCBs.

\acknowledgments

The work has been supported by the CERN Strategic Programme on Technologies for Future Experiments (\url{https://ep-rnd.web.cern.ch/}).

\end{document}